\documentclass[aps,prb,twocolumn,epsf,floatfix]{revtex4}
\usepackage{txfonts}

\usepackage{epsfig}
\usepackage{color}
\usepackage{graphicx}

\begin{document}

\title{Effects of interface resistance asymmetry on local and non-local magnetoresistance structures}

\author{Tetsufumi Tanamoto}
\affiliation{Corporate R \& D center, Toshiba Corporation,
Saiwai-ku, Kawasaki 212-8582, Japan}

\author{Hideyuki Sugiyama}
\affiliation{Corporate R \& D center, Toshiba Corporation,
Saiwai-ku, Kawasaki 212-8582, Japan}

\author{Tomoaki Inokuchi}
\affiliation{Corporate R \& D center, Toshiba Corporation,
Saiwai-ku, Kawasaki 212-8582, Japan}

\author{Mizue Ishikawa}
\affiliation{Corporate R \& D center, Toshiba Corporation,
Saiwai-ku, Kawasaki 212-8582, Japan}

\author{Yoshiaki Saito}
\affiliation{Corporate R \& D center, Toshiba Corporation,
Saiwai-ku, Kawasaki 212-8582, Japan}

\begin{abstract}
Spin injection and detection are very sensitive to the 
interface properties between ferromagnet and semiconductor. Because the interface properties 
such as a tunneling resistance can be chosen independently between the injection and detection sides,
the magnetic transport properties are considered to depend on the asymmetry of the two interfaces.  
We theoretically investigate the effect of the asymmetric interfaces of the injection side and the 
detection side on both the local and non-local magnetoresistance measurements. 
The results show the magnetoresistance ratio of local measurement structure has its maximum at the symmetric structure,  
and the effect of the asymmetry is very weak for the non-local measurement structure. 
\end{abstract}

\maketitle
\section{Introduction}
Since the successful detection of spin accumulation signals in Si at room temperature (RT), 
electrical spin injection and detection in semiconductors (SC) have been actively 
studied~\cite{Appelbaum,Dash,Suzuki,Li,Jeon,Ando,Vera-Marun,Vera-Marun2,InokuchiJAP,Ishikawa,Saito1,Jansen,Uemura} 
for realizing spin-based MOSFETs~\cite{Tanaka,Saito2,Saito3,Tanamoto}.Up to now, spin injection and detection in Si have 
been achieved using hot-electron transport~\cite{Appelbaum}, or spin-polarized tunneling~\cite{Dash,Suzuki,Li,Jeon}. 
The electrical creation and detection of spin accumulation in n-type and p-type Si 
were recently demonstrated using FM/tunnel contacts~\cite{Dash,Suzuki,Jeon} and FM/Schottky-tunnel-barrier 
contacts~\cite{Ando} up to RT. Recently, we have also succeeded in observing spin accumulation signals 
at RT by using epitaxially formed CoFe/AlOx~\cite{InokuchiJAP,Saito1} and CoFe/MgO~\cite{Ishikawa,Saito1} electrodes on 
Si-on-insulator substrates with Si (100) surfaces. In the case of CoFe/MgO electrodes, 
we have observed relatively large spin accumulation signals with relatively long spin 
relaxation times ($\tau=1.4$ ns) at RT~\cite{Ishikawa,Saito1}. 
For ferromagnet (FM)/SC systems, in general, spin injection and detection efficiencies 
strongly depend on the spin polarization of the used FM electrode, the interfacial quality, 
and the interface resistance between FM and SC. In particular, insertion of an interface 
resistance using a tunnel barrier between FM and SC is a promising method for demonstrating 
highly efficient spin injection and detection~\cite{Schmidt,Rashba,Fert2001}. These indicate that it is better that 
the interface resistance of the injection side should be sufficiently high in order to solve 
the issue of conductance mismatch between the FM and the SC. Regarding the detection side, 
in order to prevent the refection of spin-polarized electron at the interface~\cite{Fert2001}, low interface 
resistance region might be better. Then, there arise questions such as (1) whether 
the injection resistance should be larger than detection resistance and (2) Is there 
any different correlation about the interface resistance asymmetry between local magnetoresistance (MR) 
setup and non-local MR setup?

Here we theoretically study the effects of the interface resistance asymmetry between 
FM and nonmagnetic conductor (N) on the MR in both local MR setup (Fig.1(a)) and non-local 
MR setup (Fig.1(b)). Because both the spin injection and detection affect the spin-dependent 
transport properties, the symmetry of the device structure is considered to play 
an important role for the MR ratio. We extend the standard theories by Fert {\it et al}.~\cite{Fert2001} 
for the local MR measurement and Jedema {\it et al}.~\cite{Jedema2} for  the non-local 
MR measurement, and derive the analytical formula for the asymmetric interface resistances.

In \S 2, formulation of our model is presented. 
Section 3 is devoted to the analytical and numerical calculations and discussions for the 
local measurement setup. The non-local MR measurement setup is investigated in \S 4. 
Finally, the conclusions are given in \S 5.
We expect that these novel results will open a pathway for development of next-generation spin transistors.

\section{Formulation}
The spin accumulation effect, the difference between the 
chemical potential of the up spin $\mu_+$ and that of down spin $\mu_-$ 
is described by the diffusion equations~\cite{Fert2001,ValetFert,Jaffres,Fert2012,Jedema2}.
The relationship between the spin current density $J_s$   
and $\mu_{s}$ ($s=\pm$) is given by $
J_s=\frac{\sigma_s}{e} \frac{\partial }{\partial x} \mu_s
$ where $\sigma_s$ is a spin-dependent conductivity. The diffusion equation 
for the chemical potential is given by
\begin{equation}
 \frac{\partial^2 (\mu_+-\mu_-)}{\partial x^2} =\frac{(\mu_+-\mu_-)}{l_{\eta}^2},  
\end{equation}
where 
\begin{equation}
l_{\eta}^{-2} =l_{\eta+}^{-2}+l_{\eta-}^{-2}
\end{equation}
is an average 
spin diffusion length for $F$ region ($\eta=F$) and $N$ region ($\eta=N$). 
The boundary conditions at the interface ($z=z_0$) are (1) spin current is continuous 
$J_\pm (z_0^+) =J_+ (z_0^-)$ and  (2) 
the chemical potential is continuous at the interface $z_0$ such as 
$\mu_\pm (z_0^+) -\mu_+ (z_0^-)=r_\pm J_\pm (z_0)$.
We also write $\rho_{\pm} =2[1\pm \beta]\rho_F$ and $\rho_{\pm} =2\rho_N$ for the resistivity 
of the ferromagnet $F$ and the nonmagnet $N$. 
The important quantities here are $r_F\equiv \rho_F l_F$ and $r_N\equiv \rho_N l_N$.
We consider the spin-dependent interface resistances $r_\pm^L$ (injection side) and $r_\pm^R$
(detection side) for a unit surface described by
\begin{eqnarray}
r_\pm^L &=&2r_b^L [1 \mp \gamma^L], \\
r_\pm^R &=&2r_b^R [1 \mp \gamma^R]. 
\label{def_r}
\end{eqnarray}

\begin{figure}
\centering
\includegraphics[width=6cm]{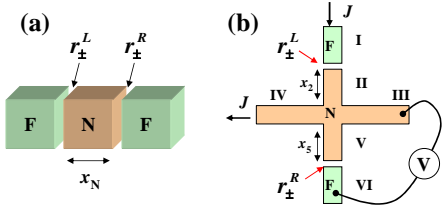}  
\caption{Two geometries for a F/N/F structure.
(a) Local measurement setup~\cite{Fert2001}. (b) non-local measurement setup\cite{Jedema2}.
$r_\pm^L$ and $r_\pm^R$ are spin ($\pm$) dependent interface tunneling resistances for a unit surface.}
\label{place0}
\end{figure}

\section{Local measurement setup}
Here we consider the effect of the asymmetric structure on the local measurement setup (Fig.1(a)). 
In this case, 
Fert {\it et al.}\cite{Fert2001} showed that there is an appropriate interface resistance condition 
of the injection side in which 
the MR ratio has a peak when there is an impedance matching $r_N\sim r_b$. 
However, because they assume the symmetric structure, it was unclear 
the appropriate condition at the detection side.
Thus we focus on the effect of the asymmetry of the device structure 
on the MR ratio.

Following Ref.\cite{Fert2001}, we assume the solutions of the diffusion equations 
for the left, middle and right electrodes. For example, 
$\bar{\mu}_{L\pm} (z) =(1-\beta^2) e\rho_F Jz +K_1^L\pm(1\pm\beta) \Delta \mu^L (z)$
and $
\Delta \mu^L (z) =K_2^L e^{z/l_F} +K_3^L e^{-z/l_F}$, 
where $\beta$ and $J$ are a bulk spin asymmetry coefficient and 
the current that flows through the system, respectively. $K_i^L$ ($i=1,2,3$) is an unknown coefficient
to be determined by the boundary conditions. 
By the same procedure as that of Ref.~\cite{Fert2001}, we obtain the resistance of anitiparallel (AP) 
configuration, $r_{AP}$, and 
that of parallel (P) configuration,
$r_{P}$. The resistance change $\Delta R=r_{AP}-r_{P}$ for Fig.1(a) is given by 
\begin{eqnarray}
\Delta R 
\!\!\! &=&\!\! \frac{2 \gamma_L \gamma_R r_b^Lr_b^R + 3\beta r_F[\gamma_L r_b^L+\gamma_R r_b^R] + 
   4 \beta^2 r_F^2}{\Theta_0},
\end{eqnarray}
where 
\begin{eqnarray}
\lefteqn{ \Theta_0  \equiv  \left(r_F+ (r_b^L+r_b^R)/2 \right) \cosh (x_N/l_N) }
\nonumber \\
&+&(r_N/2) \left[  1+ (r_F +r_b^L)(r_F + r_b^R)/r_N^2 \right]\sinh (x_N/l_N).
\label{DeltaR}
\end{eqnarray} 
Similarly, for a multilayer structure as Ref.\cite{Fert2001}, we obtain
\begin{equation}
\Delta R = \frac{2(\gamma_L r_b^L + \beta r_F )(\gamma_R r_b^R + \beta r_F)}{
\Theta_1
},
\label{FertEq}
\end{equation}
where 
\begin{eqnarray}
\lefteqn{ \Theta_1=(r_F+ (r_b^L+r_b^R)/2) \cosh (x_N/l_N) }
\nonumber \\
&+&(r_N/2) [  1+ (r_F +r_b^L)(r_F + r_b^R)/r_N^2]\sinh (x_N/l_N).
\label{FertEq2}
\end{eqnarray} 
In Eqs.(\ref{FertEq}) and (\ref{FertEq2}), we assume that thickness of $F$ is much larger 
than $l_F$ as in Ref.\cite{Fert2001}.   
The MR ratio is defined by $(r_{AP}-r_{P})/r_{P}$. 
Fig.2 shows the numerical result of Eq.(\ref{FertEq}) 
as a function of $r_b$ and $\delta_r \equiv (r_b^R-r_b^L)/r_b$
for the case of 
$\gamma_L=\gamma_R$ ((a) and (b)) and that 
in which $\gamma_L$ and $\gamma_R$ change as  
$\gamma_R=\gamma_0(1+\delta_r/2)$ and $\gamma_L=\gamma_0(1-\delta_r/2)$ ((c)). 
For $\gamma^L=\gamma^R$(Fig.2~(a)(b)), the MR ratio has its maximum when interface resistance is 
symmetric ($r_b^L=r_b^R$). We can also see that the effect of the asymmetry on the MR ratio 
is not large as long as the difference between $r_b^L$ and $r_b^R$ is not large.
Thus we can say that the impedance matching condition is also satisfied 
at the detection side.
Fig.2~(c) shows the case that $\gamma_L$ and $\gamma_R$ change with $r_b^L$ and $r_b^R$.
It can be found that the MR ratio increases as the asymmetry increases.
Note that the minimum structure at the symmetric point $r_b^L=r_b^R$ can be seen 
at $\gamma_0 \gtrsim 0.7$. Thus, this result says that the strong interface 
asymmetry coefficient $\gamma$ changes the magnetic properties of the system.

Let us consider the effect of the asymmetry around the symmetric point analytically based on Eqs.(\ref{DeltaR}) and (\ref{FertEq}). 
We consider the impedance matching region in which $r_b \sim r_N \gg r_F$. Then $r_F$ can be neglected.
In this region Eqs.(\ref{DeltaR}) and (\ref{FertEq}) have the same form given by
\begin{equation}
\Delta R = \frac{2 \gamma^2 (1-\delta_r^2) r_b^2 (1-\delta_b^2) }{
    r_b \cosh \frac{t^N}{l^N} 
          +\frac{r_N}{2} \left[  1+ \frac{r_b^2 (1-\delta_b^2)}{r_N^2} \right]\sinh \frac{t^N}{l^N}  
},
\end{equation}
where
\begin{eqnarray}
r_b^L \!\!\!&=&\!\!\!\! r_b (1+\delta_b), \ \  r_m^L =r_b (1+\delta_b) \gamma (1+\delta_r), \label{delta_b1}\\
r_b^R \!\!\!&=&\!\!\!\! r_b (1-\delta_b), \ \  r_m^R =r_b (1-\delta_b) \gamma (1-\delta_r),
\label{delta_b2}
\end{eqnarray}
assuming $\delta_b \ll 1$ and $\delta_r \ll 1$. 
In particular, when $r_N=r_b$, we have
\begin{equation}
\Delta R = \frac{2 \gamma^2 (1-\delta_r^2) r_b (1-\delta_b^2) }{
     \cosh \frac{t^N}{l^N} 
          +\frac{2-\delta_b^2}{2} \sinh \frac{t^N}{l^N}  
}.
\end{equation}

Thus, $\Delta R$ has its maximum value at its symmetric point ($\delta_b=0=\delta_r$). 

\begin{figure}
\centering
\includegraphics[width=7cm]{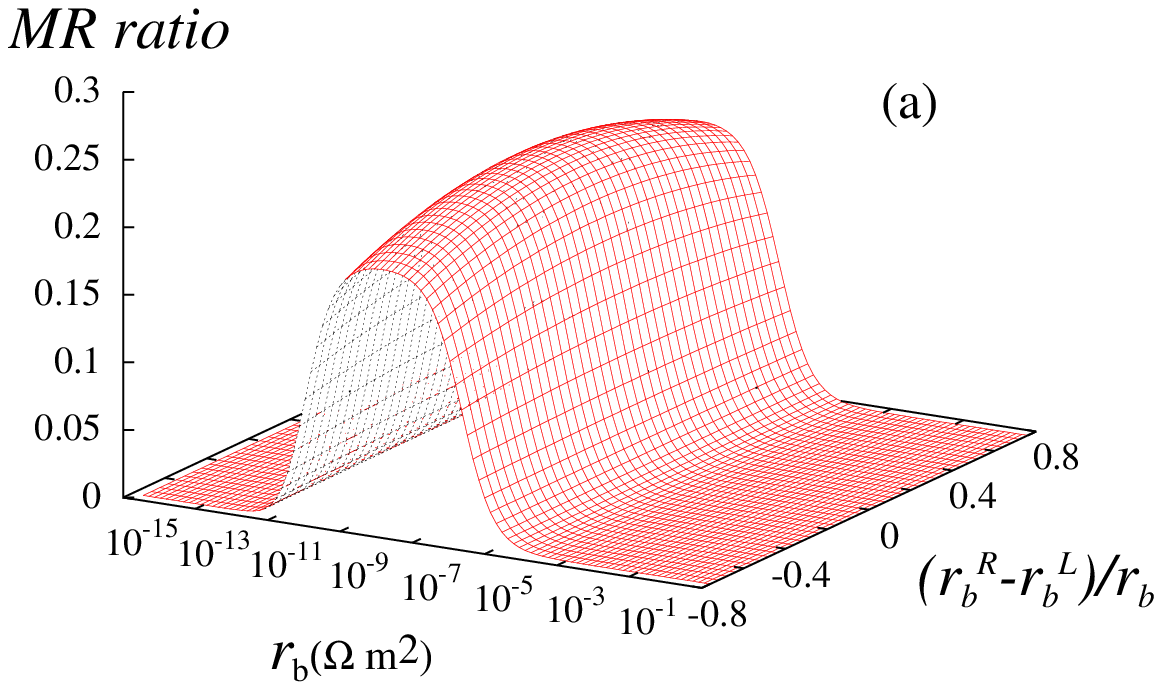}  
\includegraphics[width=7cm]{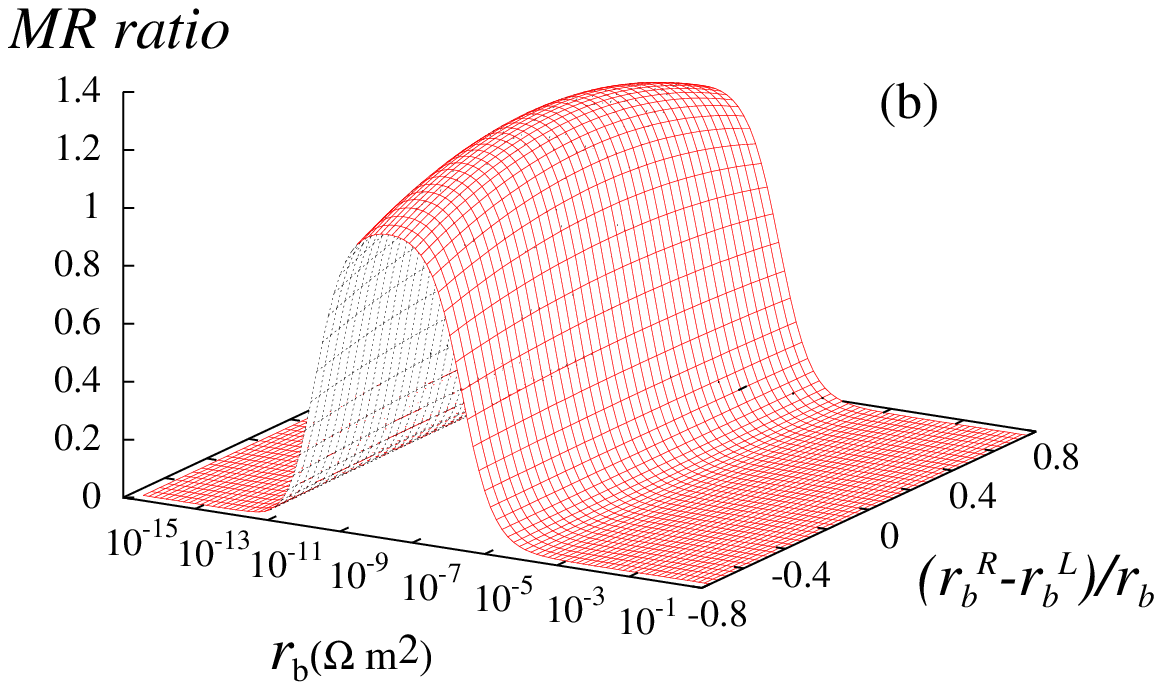}  
\includegraphics[width=7cm]{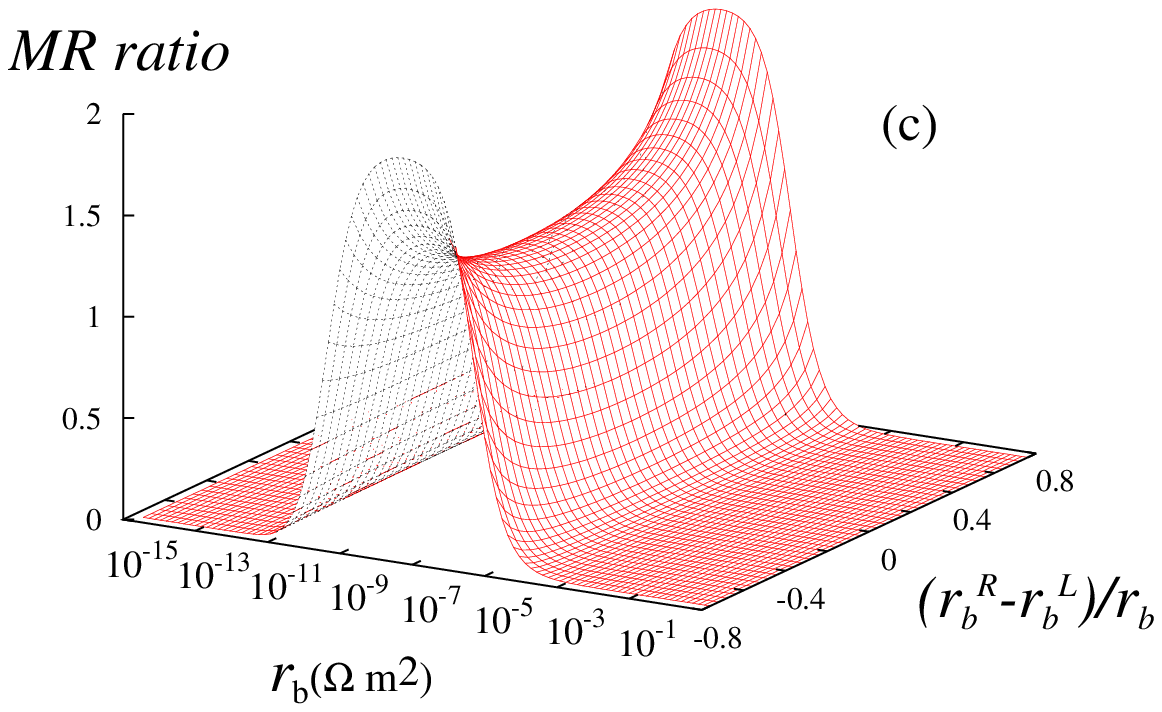}
\caption{MR ratios of the local measurement setup (Fig.1(a)) as functions of 
the average interface resistance $r_b^R$ and the asymmetry $r_b^L/r_b^R$. 
(a) $\gamma_L=\gamma_R=0.5$. 
(b) $\gamma_L=\gamma_R=0.8$.  $x_N=250$nm, 
(c) $\gamma_R=\gamma_0 (1+\delta_r/2)$ and $\gamma_L=\gamma_0 (1-\delta_r/2)$ with $\gamma_0=0.8$.
$l_N=1\mu$m, $r_N= 4.0  \times 10^{-9} \Omega m^2$, 
$l_F=5$ nm, $b=0.46$, $r_F=4.5\times 10^{-15} \Omega m^2$. In (a), $r_b^L/r_b^R=1$ corresponds to Fig.3 of Ref.\cite{ValetFert}.
}
\label{place}
\end{figure}
\section{Non-local measurement setup}
Here we consider the effect of the asymmetric structure on the non-local measurement setup (Fig.1(b)). 
Jedema {\it et al.}~\cite{Jedema2} showed the case of no interface resistances in Fig.1(b).
In case of the non-local measurement setup, we can also introduce the asymmetry of $x_2$ and $x_5$. 
We can also investigate the effect of distance between the voltage part and the current part.
Here, the diffusion equations that we use here are the same as those in Ref.~\cite{Jedema2}:
\begin{eqnarray}
\mu_{\pm}^{\rm (I)}&=& A-\frac{Je}{\sigma_F} x \pm \frac{2C}{\sigma_F(1\pm\beta)} e^{-x/l_F}, 
 \\
\mu_{\pm}^{\rm (II)}&=& -\frac{Je}{\sigma_N}x \pm \frac{2}{\sigma_N} 
[E e^{-x/l_N}+F e^{x/l_N} ], 
 \\
\mu_{\pm}^{\rm (III)}&=&\! \pm \frac{2G}{\sigma_N} e^{-x/l_N}, 
 \\
\mu_{\pm}^{\rm (IV)}&=&\! \frac{Je}{\sigma_N}x \pm \frac{2G}{\sigma_N}e^{-x/l_N}, 
 \\
\mu_{\pm}^{\rm (V)}&=& \!\pm \frac{2}{\sigma_N }[H e^{-x/l_N}+Ke^{x/l_N} ],
 \\
\mu_{\pm}^{\rm (VI)}&=& \!B \pm \frac{2D}{\sigma_F(1\pm\beta)}e^{-x/l_F},
\end{eqnarray}
where $A$, $B$, $C$, $D$, $E$, $F$, $H$, $K$ and $G$ are unknown constants that 
are determined by boundary conditions. 
Regions (I) and (VI) in Fig.1(b) correspond to the injection and detection ferromagnetic contacts,
respectively.
Regions (II)-(V) correspond to the intermediate semiconductor regions.
The electric current flows from the region (I) to the region (IV). 
In regions (III) and (V), spin polarized current flows while the electric current 
does not flow. Although in the general non-local measurement, there is a finite distance between the 
region (II)-(IV) and the region (III)-(V), as long as we consider the magnetic properties, 
the interface between the semiconductor and the ferromagnetic contacts play the intrinsic role.
Therefore we apply the cross structure depicted in Fig.1(b). 
   
The resistance change, $\Delta R=r_{AP}-r_{P}=-2B/(eJS)$, where $S$ is the 
cross-sectional area of the nonmagnetic strip, is obtained by
\begin{eqnarray}
\Delta R \!\!&\!\!=\!\!&\!\! \frac{r_N}{S} 
\frac{4[r_F \beta  + (1-\beta^2) r_m^R][r_F \beta  + (1-\beta^2) r_m^L]}
{ 2 B_+^RB_+^L b_2b_5-B_+^RB_-^Lb_5/b_2-B_+^L B_-^R b_2/b_5 }, 
\label{DRj}
\end{eqnarray}
where $b_2=e^{x_2/l_N}$, $b_5=e^{x_5/l_N}$, $r_N=el_N/\sigma_N$,
 $r_F=el_F/\sigma_F$, $r_m^\alpha=r_b^\alpha\gamma_\alpha$, and
\begin{eqnarray}
B_\pm^L & \equiv& [r_N \pm r_b^L ](1-\beta^2) \pm r_F, \\ 
B_\pm^R & \equiv& [r_N \pm r_b^R ](1-\beta^2) \pm r_F, 
\end{eqnarray}
with $\alpha=L,R$. 
For the symmetric case, this equation is reduced to that of Ref.~\cite{Jedema2}.

The spin current polarization at the interface of the current injecting contact is 
given by $P=(J_+-J_-)/(J_+ + J_-)$ and we have
\begin{equation}
P= \frac{[(1+2b_2^2 ) B_+^R b_5/b_2  -B_-^R b_2/b_5][r_F \beta  + (1-\beta^2) r_m^L]}
{ 2 B_+^RB_+^L b_2b_5-B_+^RB_-^Lb_5/b_2-B_+^L B_-^R b_2/b_5}. 
\label{P}
\end{equation}
When $r_b=0$, this result coincides with Eq.(15) in Ref.\cite{Jedema2}.

\begin{figure}
\centering
\includegraphics[width=7cm]{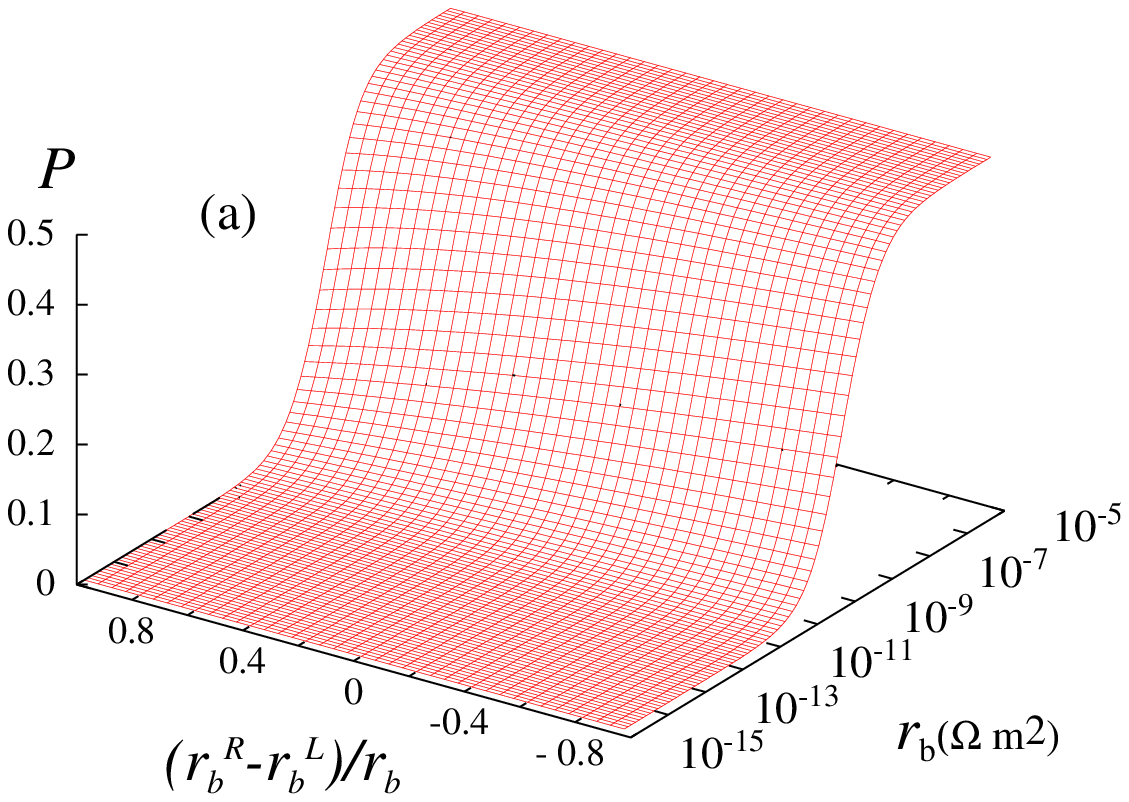}  
\includegraphics[width=7cm]{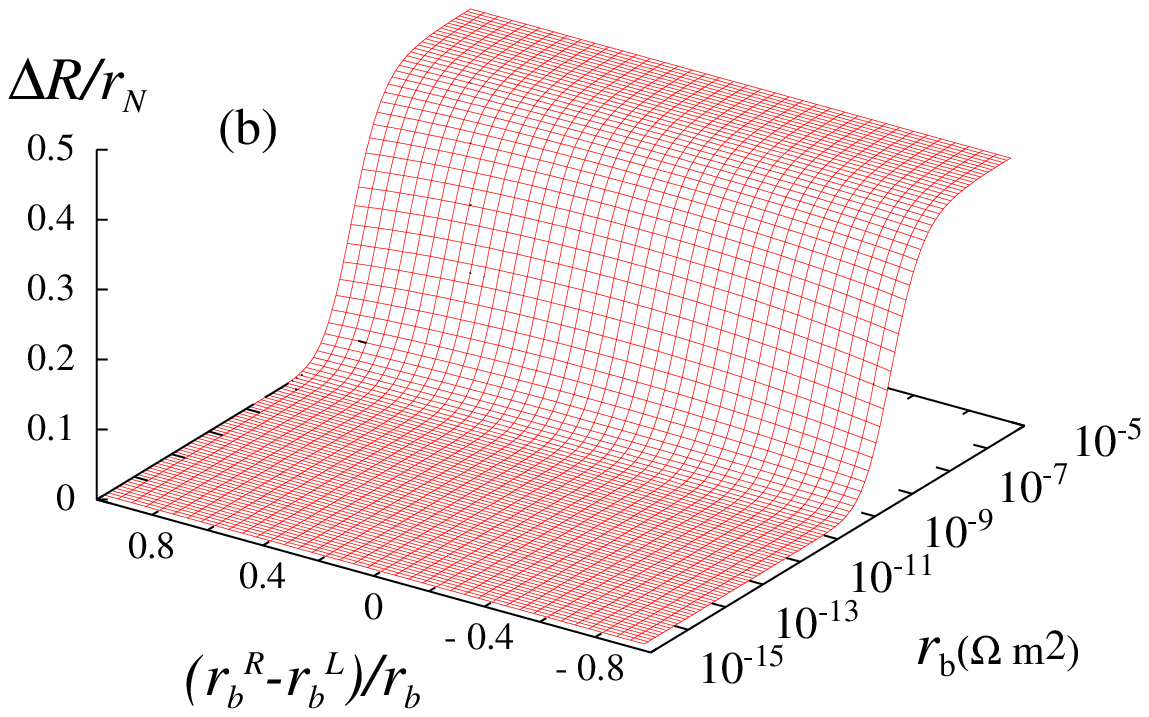}  
\caption{
(a) Spin current polarization at the interface of injection (region I) and (b) spin dependent 
resistance at the interface of detection (region VI) for the non-local measurement setup (Fig.1(b)). 
$x_2=x_5=250$nm, $l_N=1\mu$m, $r_N= 4.0\times10^{-9} \Omega{\rm m}^2$, $l_F=5$nm, $b=0.46$, 
$r_F=4.5\times 10^{-15}\Omega{\rm m}^2$. 
In (b), cross sectional area $S=1$ is assumed.
}
\label{place2}
\end{figure}

Fig.3 shows the calculated $P$ ((a)) and $\Delta R$((b)). 
Compared with Ref.\cite{Jedema2}, we can see 
that the interface resistance Eq.(\ref{def_r}) enhances $P$ and $\Delta R$
such that $P$ and $\Delta R$ are saturated as $r_b$ increases~\cite{Fukuma}.
As can be seen from these figures, the effect of the asymmetry of the left and the right tunneling 
seems to be small. 
Now let us analyze the $\Delta R$ and $P$ around their symmetric point.
Because Eq.(\ref{DRj}) is symmetric regarding the $r_b^L$ and $r_b^R$,  
it is easily inferred that the symmetric point $r_b^L=r_b^R$ will 
correspond to a maximum or minimum point similar to 
the local measurement case (Fig.1(a)) when $x_2=x_5$.
Because $r_N \gg r_F$, we can neglect $r_F$ in Eq.(\ref{DRj}). 
By using Eqs.(\ref{delta_b1}) and (\ref{delta_b2}), for $r_b \gg r_N$, we have
\begin{equation}
\Delta R  
%
\approx 
\frac{2 r_N  \gamma^2 (1-\delta_r^2) }
{ S(b_5\cosh \frac{x_2}{l_N}+b_2\cosh \frac{x_5}{l_N})  }.
\end{equation}
When $r_b \ll r_N$
\begin{equation}
\Delta R  
\approx 
\frac{2 r_b^2 \gamma^2 (1-\delta_b^2)(1-\delta_r^2) }
{ Sr_N ( b_5\sinh \frac{x_2}{l_N} 
       +  b_2\sinh \frac{x_5}{l_N} ) }.
\end{equation}
These quantities have their maximum values at $\delta_b \sim 0$ and $\delta_r \sim 0$. 
Thus, the MR is found to have their maximum at the symmetric point.
Similarly when $r_N \gg r_b$, we have
\begin{eqnarray}
P &\rightarrow & \frac{  b_2b_5+ \sinh \frac{x_5-x_2}{l_N}  }
{   b_5 \sinh \frac{x_2}{l_N}  + b_2 \sinh \frac{x_5}{l_N}  } 
\frac{r_m^L}{r_N}
\nonumber \\
&=& \left[1 +\frac{2}{2b_2^2 (2-1/b_5^2)-1}\right]\frac{r_m^L}{r_N}.
\end{eqnarray}
Thus, in this region, as $b_2$ decreases ($x_2$ decreases) and 
$b_5$ increases ($x_5$ increases), $P$ increases. 
However, because  $r_N \gg r_b$, the absolute value of $P$ is small 
such as $P \sim {r_m^L}/{r_N} \ll 1$.
When $r_N \ll r_b$ 
\begin{eqnarray}
P &\rightarrow & \frac{  b_2b_5 + \cosh \frac{x_5-x_2}{l_N} }
{   b_5\cosh \frac{x_2}{l_N}  
 +  b_2\cosh \frac{x_5}{l_N} }\frac{ r_m^L}{ r_b^L}=\gamma.
\end{eqnarray}
Thus, for $r_N \ll r_b$, $P$ is determined only by $\gamma$.
The effect of the asymmetry can be neglected in this region.
Thus, in contrast with the local measurement setup, we can see the 
effect of asymmetry only in the region $r_N \gg r_b$. 

Next we investigate in what region the asymmetry affects the spin polarized current. 
Fig.4 shows $P$ for $r_b^L/r_b^R \neq 1$ when $x_2$ changes. 
We can see that the large asymmetry shifts the impedance matching points.
In addition, $P$ for $r_b^R<r_b^L $ ($r_b^L=2r_b^R$) is better than that 
for $r_b^R >r_b^L $ ($r_b^L=0.05r_b^R$). 
This result shows that, as noted in the Introduction, 
low interface resistance is better in the detection side ($R$).
Fig.\ref{place4} (a) shows the results when the length of the nonmagnet is as short 
as the spin diffusion length such as $x_2=x_5=8$nm and $l_F=5$nm.
We can see that the effect of 
the asymmetry becomes remarkable for $r_b <r_N$. 
This is because the 
distance between the left and the right junctions is small and the effect of the two 
junctions is sufficiently strong. 
On the contrary, the resistance change $\Delta R/r_N$ (Fig.\ref{place4} (b)) 
again does not strongly depend on the interface asymmetry.

\begin{figure}
\centering
\includegraphics[width=7cm]{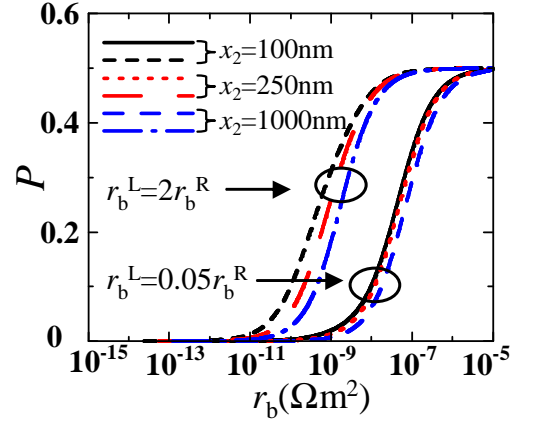}  
\caption{
Spin current polarizations for $r_b^L/r_b^R\neq 1$. $x_2=x_5$ when $x_2$ changes 
for the non-local measurement setup (Fig.1(b)).
}
\label{place3}
\end{figure}

\begin{figure}
\centering
\includegraphics[width=7cm]{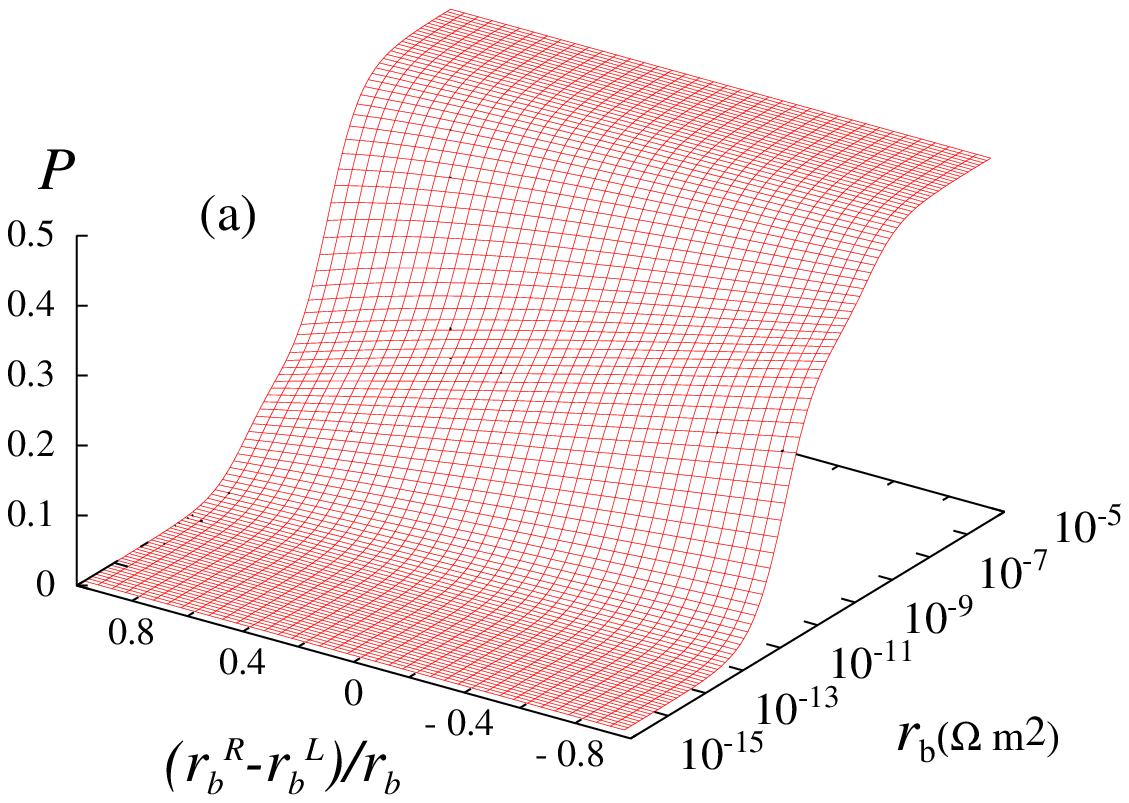}  
\includegraphics[width=7cm]{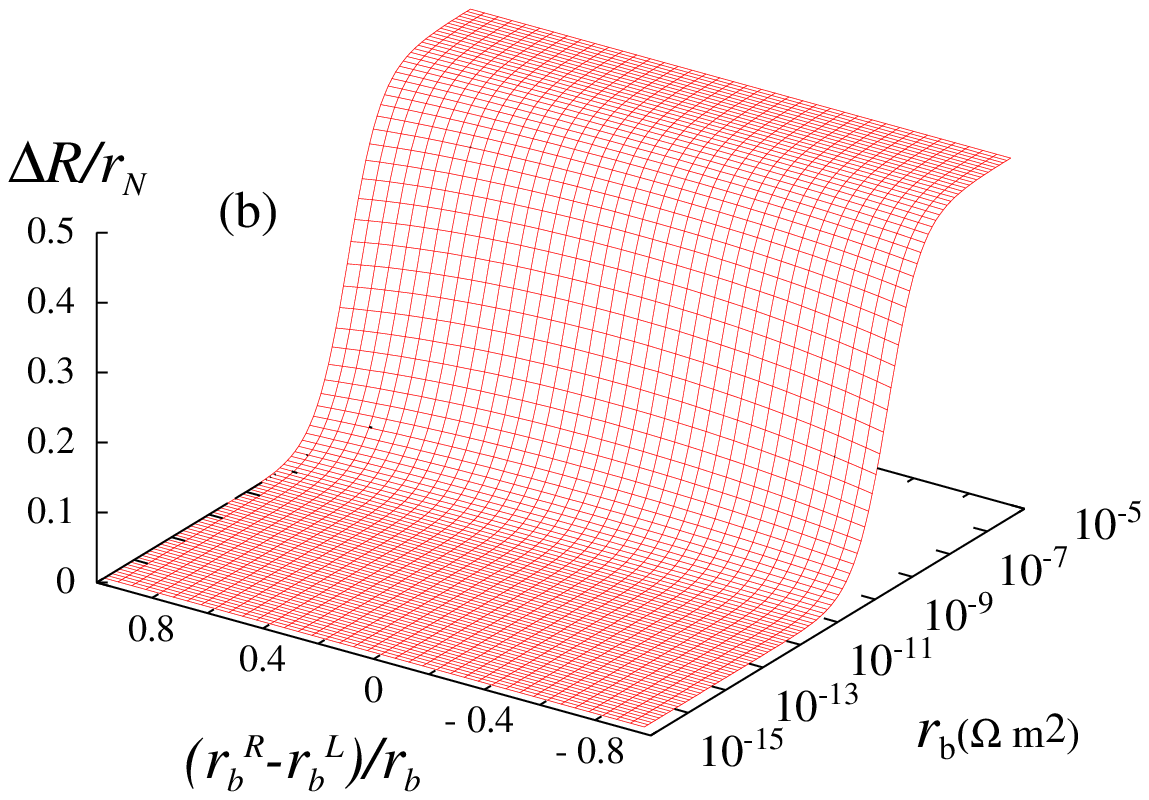}  
\caption{
(a) Spin current polarization at the interface of injection (region I) and (b) spin dependent 
resistance at the interface of detection (region VI) for the non-local measurement setup (Fig.1(b)). 
$x_2=x_5=8$nm, $l_N=1mu$m, $r_N= 4.0\times10^{-9} \Omega{\rm m}^2$, $l_F=5$nm, $b=0.46$, 
$r_F=4.5\times 10^{-15}\Omega{\rm m}^2$. 
In (b), cross sectional area $S=1$ is assumed.
}
\label{place4}
\end{figure}

\section{Conclusions}
We studied the effects of the interface  asymmetry
between ferromagnet and nonmagnet on the spin-dependent transport properties  
for both the local and non-local measurement setups, based on diffusion equations.
We showed that the MR ratio has its maximum value at the symmetric structure 
of the conventional case of $\gamma_L=\gamma_R$ 
for the local measurement setup. For the non-local measurement setup, 
both the resistance change and the spin current polarization monotonically increase and  
saturate as the interface resistance increases, even if there is the interface asymmetry. 
In particular the resistance change and the spin current polarization 
is affected by the interface asymmetry only below the impedance matching region.  
We can conclude that, as long as the asymmetry is not large, spin-dependent transport properties 
are not strongly affected by the asymmetry of the interface resistance 
for the non-local measurement setups.

\acknowledgements
We thank A. Nishiyama, J. Koga and S. Fujita for useful discussions.



\begin{thebibliography}{99}
\bibitem{Appelbaum}
I. Appelbaum, B. Huang, and D.J. Monsma: Nature (London) {\bf 447} (2007) 295. 

\bibitem{Dash} 
S.P. Dash, S. Sharma, R.S. Patel, M.P. Jong, and R. Jansen: Nature (London) {\bf 462} (2009)491.

\bibitem{Suzuki} 
T. Suzuki, T. Sasaki, T. Oikawa, M. Shiraishi, Y. Suzuki, and K. Noguchi:
Appl. Phys. Express {\bf 4} (2011) 023003.

\bibitem{Li} 
C.H. Li, O.M. J. van't Erve, and B.T. Jonker: Nature Commun. {\bf 2} (2011) 245.

\bibitem{Jeon}  
K.R. Jeon, B.C. Min, I.J. Shin, C.Y. Park, H.S. Lee, Y.H. Jo, and S.C. Shin: Appl. Phys. Lett. {\bf 98} (2011) 262102.

\bibitem{Ando} 
Y. Ando, K. Kasahara, S. Yamada, Y. Maeda, K. Masaki, Y. Hoshi, 
K. Sawano, M. Miyao, and K. Hamaya: Phys. Rev. B {\bf 85} (2012) 035320.

\bibitem{Vera-Marun}
I. J. Vera-Marun, V. Ranjan, and B. J. van Wees: Nature Phys.
{\bf 8} (2012) 313.

\bibitem{Vera-Marun2}
M. H. D. Guimaraes, A. Veligura, P. J. Zomer, T. Maassen,
I. J. Vera-Marun, N. Tombros, and B. J. van Wees: Nano Lett.
Article ASAP, DOI 10.1021/nl301050a

\bibitem{InokuchiJAP} 
T. Inokuchi, M. Ishikawa, H. Sugiyama, Y. Saito, N. Tezuka: J. Appl. Phys. {\bf 111} (2012) 07C316.

\bibitem{Ishikawa} 
M. Ishikawa, H. Sugiyama, T. Inokuchi, K. Hamaya and Y. Saito: Appl. Phys. Lett.{\bf 100} (2012) 252404.

\bibitem{Saito1} 
Y. Saito, M. Ishikawa, T. Inokuchi, H. Sugiyama, T. Tanamoto, K. Hamaya, N. Tezuka: To be published in IEEE Tran. Magn.

\bibitem{Jansen}
R. Jansen: Nature. Mater. {\bf 11} (2012) 400.

\bibitem{Uemura}
T. Uemura,  T. Akiho, M. Harada, K.-i. Matsuda, and M. Yamamoto:
Appl. Phys. Lett. {\bf 99} (2011) 082108.


\bibitem{Tanaka} 
S. Sugahara and M. Tanka: Appl. Phys. Lett. {\bf 84} (2004) 2307.

\bibitem{Saito2} 
Y. Saito, T. Marukame, T. Inokuchi, M. Ishikawa, H. Sugiyama, and T. Tanamoto, Thin Solid Films {\bf 519} (2011)8266.
 
\bibitem{Saito3} 
Y. Saito, T. Inokuchi, M. Ishikawa, H. Sugiyama, T. Marukame, and T. Tanamoto, J. Electrochem. Soc. {\bf 158}(2011)H1068.

\bibitem{Tanamoto}
T. Tanamoto, H. Sugiyama, T. Inokuchi, T. Marukame, M. Ishikawa, K. Ikegami, and Y. Saito: 
J. Appl. Phys. {\bf 109} (2011) 07C312.

\bibitem{Schmidt} 
G. Schmidt, D. Ferrand, L. W. Molenkamp, A. T. Filip, and B. J.
van Wees: Phys. Rev. B {\bf 62} (2000) R4790.

\bibitem{Rashba} 
E.I. Rashba: Phys. Rev. B {\bf 62} (2000) R16267.

\bibitem{Fert2001} 
A. Fert and H. Jaffr'es: Phys. Rev. B {\bf 64} (2001)184420.







\bibitem{ValetFert}
T. Valet and A. Fert: Phys. Rev. B {\bf 48} (1993) 7099.

\bibitem{Jaffres}
H. Jaffr`es, J.-M. George, and A. Fert: Phys. Rev. B {\bf 82} (2010) 140408(R).

\bibitem{Fert2012}
P. Laczkowski, L. Vila, V.-D. Nguyen, A. Marty, J.-P. Attane, H. Jaffres, J.-M. George, and A. Fert:
Phys. Rev. B {\bf 85} (2012) 220404(R). 

\bibitem{Jedema2}
F.J. Jedema, M.S. Nijboer, A.T. Filip, and B.J. van Wees:
Phys. Rev. B {\bf 67} (2003) 085319.

\bibitem{Fukuma}
Y. Fukuma, L. Wang, H. Idzuchi, and Y. Otani: Appl. Phys. Lett. {\bf 97} (2010) 012507.


\end{thebibliography}
\end{document}